\documentclass[twocolumn,secnumarabic,amssymb,showpacs,nobibnotes,aps,prl]{revtex4}
\usepackage{amsmath}
\usepackage{latexsym}
\usepackage{float}
\usepackage{amssymb}
\usepackage{graphicx}
\usepackage{dcolumn}
\def\ba{\begin{eqnarray}}
\def\ea{\end{eqnarray}}
\def\be{\begin{equation}}
\def\ee{\end{equation}}

\def\bm{\begin{math}}
\def\me{\end{math}}

\newcommand{\dummy}

\begin{document}
%\widetext

\preprint{APS/123-QED}

\title{Critical Dynamics in a Binary Fluid: Simulations and Finite-size Scaling}

\author{Subir K. Das,$^{1}$ Michael E. Fisher,$^1$ Jan V. Sengers,$^1$ J\"urgen Horbach,$^2$ and Kurt Binder$^2$}

\affiliation{$^1$Institute~for~Physical~Science~and~Technology,~University~of~Maryland,~College~Park,~MD~20742,~USA\\
$^2$Institut f\"{u}r Physik, Johannes
Gutenberg-Universit\"{a}t, D-55099 Mainz, Staudinger Weg 7, Germany}
\date{\today}

\begin{abstract}
We report comprehensive simulations of the critical dynamics of a symmetric binary
Lennard-Jones mixture near its consolute point. The self-diffusion coefficient 
exhibits no detectable anomaly. The data for the shear viscosity 
and the mutual-diffusion coefficient are fully consistent with the
asymptotic power laws and amplitudes predicted by renormalization-group and mode-coupling theories
{\it provided} finite-size effects and the background contribution 
to the relevant Onsager coefficient are suitably accounted for. This resolves a
controversy raised by recent molecular simulations.
\end{abstract}

\pacs{64.60.Ht, 64.70.Ja}

\maketitle

\begin{figure}[htb]
\centering
\includegraphics*[width=0.47\textwidth]{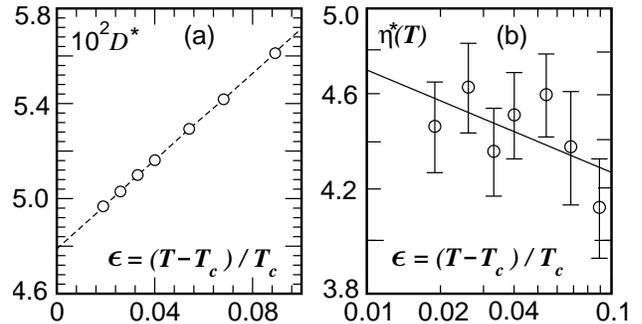}
\caption{\label{fig1}(a) Reduced self-diffusion constant $D^*$
for a system of $N=6400$ particles as
a function of $T$.
The dashed line guides the eye and shows that $D(T_c)$
is nonzero.
(b) Log-log plot of the reduced shear viscosity for the same system.}
\end{figure}

{\bf Introduction}.---\,Thermodynamic~~ and ~~transport\\ properties exhibit critical-point singularities 
with exponent values and amplitude ratios
that are the same for systems belonging to the same universality class.
As regards static critical behavior, it has been well established that
fluids of molecules with short-range interactions belong to the
universality class of three-dimensional Ising-type systems \cite{1}. It is expected that
the dynamic critical behavior of fluids conforms to that of model $H$ in the
nomenclature of Hohenberg and Halperin \cite{2}. Recently, there has been a revival of
interest in critical phenomena, one reason being that
computer-simulation techniques have matured sufficiently  
that they can provide interesting detailed
information concerning the static critical behavior \cite{3,4,5}.
For instance, recent Monte Carlo simulations have 
provided new insights concerning the nature of the
scaling fields in asymmetric fluids \cite{5}.\\
\indent The status of computer simulations of the dynamic critical behavior of
fluids is much less satisfactory. Specifically, on the basis of a recent 
molecular dynamics simulation
of a binary fluid Jagannathan and Yethiraj (JY) \cite{7} 
concluded that the dynamic exponent $x_D$ that governs the 
slowing down of critical fluctuations
differs substantially from the value predicted by
renormalization-group  or mode-coupling theory \cite{2,nn7}. 
Sengers and Moldover have pointed out that the
conclusion of JY is also in disagreement with 
reliable experimental evidence \cite{8}.\\
\indent To address this issue we have undertaken a comprehensive study of 
the dynamic critical behavior of a symmetric Lennard-Jones mixture (A+B)
near its consolute point. We find that the data for the
transport property that determines the nature of 
critical slowing-down are significantly affected by finite-size
effects and by a `background' contribution arising from
fluctuations at small length scales. After properly accounting for both these effects
our extensive simulations of the critical
dynamics are fully consistent, both with current theoretical
predictions and with the best available experimental evidence.\\
\indent {\bf The model}.---~Starting from the standard~~($12$,\,$6$) Lennard-Jones
potential with parameters $\varepsilon_{\alpha\beta}$
and $\sigma_{\alpha\beta}$ ($\alpha$,\,$\beta$\,=\,A,\,B)
we construct a truncated potential 
which is strictly zero for $r$\,$>$\,$r_c$ and
makes both the potential and the force continuous at $r$\,=\,$r_c$ \cite{new9}.~For
the parameters, we take
$\sigma_{AA}$ =\,$\sigma_{BB}$\,=\,$\sigma_{AB}$\,=\,$\sigma$,~
$\varepsilon_{AA}$\,=\,$\varepsilon_{BB}$\,=\,$2\varepsilon_{AB}$\,=\,$\varepsilon$,~$r_c$\,=\,$2.5\sigma$,  and 
define the reduced temperature as $T^*$\,=\,$k_{\rm B}T/\varepsilon$.
The total particle number
$N$\,=\,$N_{\rm A}$+$N_{\rm B}$
and the volume $V$\,=\,$L^3$ are chosen
so that the reduced density $\rho^*$\,=\,$\rho\sigma^3$\,=\,$N/V$ is unity and the
simulation box edge is $L/\sigma$\,$\equiv$\,$L^*$\,=\,$N^{1/3}$. For these parameters
the system is far from solid-liquid and liquid-gas transitions in the
temperature regime of interest here.\\
\indent In order to evaluate the results of computer simulations of dynamic
critical behavior, accurate information regarding the static critical 
behavior is a prerequisite. We have obtained this by using a 
semi-grandcanonical Monte Carlo (SGMC) approach \cite{new9,9,12}.
In the SGMC method, in addition to displacement moves, the
particles may switch identity (A$\rightarrow$B$\rightarrow$A)
with both the energy change
$\Delta E$ and the chemical potential difference $\Delta\mu$\,=\,$\mu_A$$-$$\mu_B$ 
entering the Boltzmann factor. 
For the case $\Delta\mu=0$, of interest here, one has 
$\langle x_{\rm A}\rangle$\,=\,$\langle x_{\rm B} \rangle$\,=$1/2$ 
(with $x_{\alpha}$\,=\,$N_{\alpha}/N$) for
$T>T_c$.\\
\indent {\bf Static Properties}.---~Since our focus is on the
dynamic critical behavior, we simply state the results found for the static 
behavior \cite{new9}. An accurate, unbiased estimate for the ~reduced critical 
temperature was obtained by plotting the fourth-order cummulant~~
$U_L=\langle (x_{\rm A}-1/2)^4\rangle_L/\langle (x_{\rm A}-1/2)^2\rangle^2_L$
as a function of $T$ for various box sizes $L$ and identifying 
$T_c$ from the asymptotically common intersection point \cite{5,new9,13}: this 
yields $T_c^*=1.4230\pm0.0005$ \cite{new9}. The concentration susceptibility
$\chi(T)$ was calculated from
$k_BT\chi=\chi^*T^*=N(\langle x_{\rm A}^2\rangle -\langle x_{\rm A} \rangle ^2)$
($T$\,$>$\,$T_c$). The correlation length $\xi(T)$ was determined from Ornstein-Zernike
plots of the Fourier transform of the concentration-concentration
correlation function, $S_{cc}(q)=T^*\chi^*/[1+q^2\xi^2+...~]$. In the
thermodynamic limit (i.e., in the absence of finite-size effects) these properties diverge
as $\chi^*\approx\Gamma_0\epsilon^{-\gamma}$ and
$\xi\approx\xi_0\epsilon^{-\nu}$ where $\epsilon=(T-T_c)/T_c$ and
we may adopt $\gamma=1.239$ and $\nu=0.629$ as the universal critical
exponents for the 3-dimensional Ising universality class \cite{14}.
Our SGMC simulations \cite{new9} then yield
$\Gamma_0=0.076\pm0.006$ and $\xi_0/\sigma=0.395\pm0.025$.\\
\indent {\bf Dynamics}.---~We investigated the dynamic behavior by implementing a microcanonical
Molecular Dynamics (MD) simulation \cite{15}. For this study,
multiple independent initial configurations
were prepared from SGMC runs with $5$$\times$$10^5$ Monte Carlo steps (MCS) per
particle. This was followed
by a microcanonical thermalization for $2$$\times$$10^5$ MD steps in the $NVT$ ensemble using the
Andersen thermostat \cite{15} before the production runs commenced.
For the MD simulations, the particle masses were taken
equal: $m_A=m_B=m$. The standard 
Verlet velocity algorithm \cite{15} was employed with a time step $\delta t^*$\,=\,$0.01/\sqrt{48}$ [in units 
$t_0=(m\sigma^2/\epsilon)^{1/2}$].\\
\indent {\bf Self-diffusion}.---\,Restricting attention to temperatures $T$\,$\geq$\,$T_c$
and to the critical concentration $x_c$\,=\,$x_{\rm A}$\,=\,$x_{\rm B}$\,=\,$1/2$,
the symmetry of our model dictates that the self-diffusion
constant is the same for A and B particles: $D_{\rm A}$\,=\,$D_{\rm B}$\,=\,$D$. We have calculated
the reduced self-diffusion coefficient $D^*$ from mean square displacements via
$(\sigma^2/t_0)D^*\equiv D=\lim_{t \rightarrow \infty} \langle [\vec{r}_{i}(0)-\vec{r}_{i}(t)]^2\rangle /6t$,
where the average $\langle\bullet\rangle$ includes all A and B particles.
The results are shown in Fig.~\ref{fig1}(a) as a function of $\epsilon$.
No anomalous critical behavior is detected and 
the linear behavior is
consistent with previous simulation studies \cite{7,17}. An MD calculation
\cite{18} has suggested a weak singularity in the
self-diffusion near vapor-liquid criticality but no corresponding anomaly
has yet been detected experimentally \cite{19}.\\
\indent {\bf Shear viscosity}.---\,The shear viscosity is expected to diverge as 
$\xi^{x_{\eta}}$ with $x_{\eta}\simeq 0.0679$
according to recent theoretical calculations \cite{20}; this
value is in good agreement with the best available experimental information \cite{8,21}.
We have calculated the reduced shear viscosity $\eta^*$ from the appropriate
Green-Kubo formula \cite{22}
\begin{equation}\label{eq7}
\eta^*=(t_0^3/\sigma Vm^2T^*)\int_0^{\infty} dt\langle 
\sigma_{xy}(0)\sigma_{xy}(t)\rangle,
\end{equation}
where the pressure tensor is $\sigma_{xy}(t)=\sum_{i=1}^N[m_iv_{ix}v_{iy}+
\frac{1}{2}\sum_{j(\neq i)}|x_i-x_j|F_y(|\vec r_i-\vec r_j|)]$ with
$\vec F$ and $\vec v_i$ the force between 
particles $i$ and $j$ and the velocity of particle $i$, respectively.
The numerical data for $\eta^*$ obtained from simulations with $N=6400$ particles
are shown in Fig.~\ref{fig1}(b).
As always in MD simulations, accurate estimation
of the shear viscosity is difficult and the $\pm5$\% error bars 
prevent us making any strong statements about the singular behavior of $\eta^*$.
But the slight increase of $\eta$ as $T\rightarrow T_c$ is
actually consistent with the 
predicted power-law divergence $\eta^*\approx \eta_0\epsilon^{-\nu x_{\eta}}$
with $\nu=0.629$ and $x_{\eta}=0.068$. The corresponding least-squares fit in Fig.~\ref{fig1}(b)
\begin{figure}[htb]
\centering
\includegraphics*[width=0.42\textwidth]{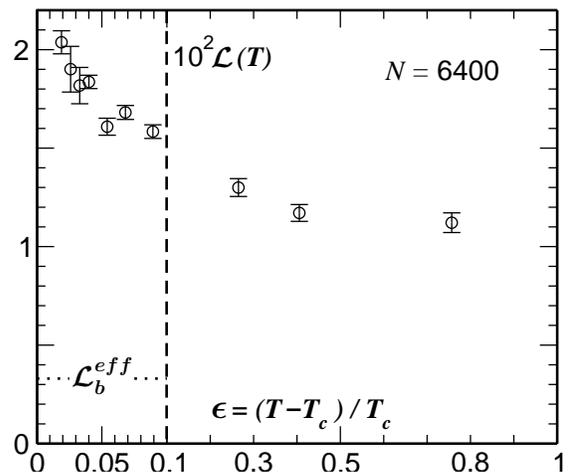}
\caption{\label{fig2} Variation with temperature of the Onsager coeffecient $\mathcal L(T)$
for a system of $N=6400$ particles.
Note the expansion of the scale for $\epsilon\leq 0.1$.
The dotted line represents an effective background contribution: see text.}
\end{figure}
yields $\eta_0=3.8_7\pm0.3$.\\
\indent {\bf Mutual diffusion}.---\,Dynamic renormalization-gro-\\up and 
mode-coupling theories predict that the mutual
diffusion coefficients $D_{{\rm AB}}(T)$ will vanish asymptotically as
$\xi^{-x_D}$, where 
\begin{equation}\label{dynamicscaling}
x_D=1+x_{\eta}\simeq 1.068,
\end{equation}
so that there is only one independent exponent characterizing the dynamic
critical behavior of fluids \cite{2}. This relation has
been verified experimentally \cite{23}.\\
\indent The mutual diffusion coefficient $D_{{\rm AB}}=(\sigma^2/t_0)D_{{\rm AB}}^*$
is related to a corresponding Onsager coefficient $\mathcal L$ via
$D_{{\rm AB}}^*=\mathcal L/\chi^*$ \cite{25}. We have calculated $D_{{\rm AB}}^*$
by adopting the result $\chi^*(T)\approx \Gamma_0\epsilon^{-\gamma}$ previously
obtained, and using MD simulations to determine 
$\mathcal L(T)$ from the appropriate Green-Kubo formula \cite{22}
\begin{equation}\label{eq8}
\mathcal L(T)=(t_0/NT^*\sigma^2)\int_0^{\infty}dt\langle
J_x^{{\rm AB}}(0)J_x^{{\rm AB}}(t)\rangle,
\end{equation}
where $\vec J^{{\rm AB}}(t)=x_{\rm B}\sum_{i=1}^{N_{\rm A}}\vec v_{i,{\rm A}}(t)-
x_{\rm A}\sum_{i=1}^{N_{\rm B}}\vec v_{i,{\rm B}}(t)$, in which
$\vec v_{i,\alpha}$ is the velocity of particle $i$ of species $\alpha$.\\
\indent If, somewhat naively, one fits the numerical values for $D_{{\rm AB}}$ 
obtained for $N=6400$ particles and $\epsilon > 0.01$ to a power law of the 
form $D_{{\rm AB}}^*\propto \xi^{-x_{{\rm eff}}}$ one finds a value of about $1.6$ 
for the effective critical exponent; this is even larger than the corresponding
value $x_{{\rm eff}}=1.26\pm0.08$ derived by Jagannathan and Yethiraj \cite{7} from
their MD simulations! Both values differ substantially from the
theoretical prediction recorded in (\ref{dynamicscaling}).\\
\indent To resolve this issue we must
focus on the Onsager coefficient $\mathcal L$ since
the simulation data for $\chi^*$ in our model
are consistent with Ising criticality \cite{new9}. While the divergence 
of $\chi^*$ near $T_c$ is strongly dominated by long-range fluctuations, it
is known that the Onsager coefficient of fluid mixtures near a consolute point (or,
its equivalent, the thermal conductivity of a fluid near a vapor-liquid
critical point) contains a critical enhancement $\Delta\mathcal L(T)$
due to long-range fluctuations {\it together} with a significant background
which arises from fluctuations at small length scales \cite{25,26} and has weak
temperature dependence \cite{27}: thus we write
\begin{equation}
\mathcal L(T)=\Delta\mathcal L(T)+\mathcal L_b(T).
\end{equation}
Such a separation has proved essential in reconciling
experimental data for $D_{{\rm AB}}(T)$ with theory \cite{27}.\\
\begin{figure}[htb]
\centering
\includegraphics*[width=0.45\textwidth]{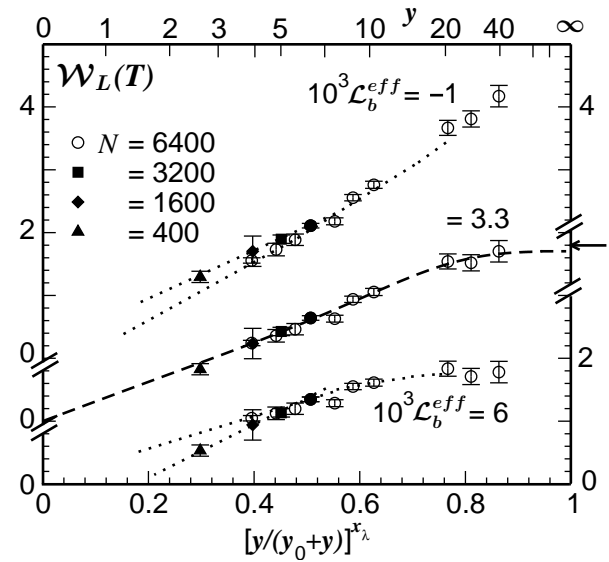}
\caption{\label{fig3} Finite-size scaling plots of the critical part of
the Onsager coefficient for trial values of the effective
background $\mathcal L_b^{eff}$ with $y=L/\xi$. We have
accepted $x_{\lambda}=0.90$ and set $y_0=7$.
The filled symbols represent data for different system
sizes at $T^*=1.48$. The arrow on the right marks the
theoretical value (\ref{qvalue}) for the critical
amplitude $Q$. The dotted lines guide the eye; the
dashed line is a scaling-function fit embodying the optimal value of $\mathcal L_b^{eff}$: see text.}
\end{figure}
\indent In Fig.~\ref{fig2} we show a plot vs. $\epsilon$ of the numerical data
obtained for $\mathcal L$ from the simulations with $N=6400$ particles.
The data do indeed suggest the presence
of a significant background. Theory predicts that $\Delta\mathcal L$
satisfies a Stokes-Einstein relation of the form
$\Delta \mathcal L=R_DT^*\chi^*\sigma/6\pi\eta^*\xi$, where
$R_D$ is a universal dynamic amplitude ratio that is of order
unity \cite{2,26}. It thus follows that $\Delta\mathcal L$ should diverge as
\begin{equation}\label{powerlawons}
\Delta\mathcal L\approx QT^*\epsilon^{-\nu_{\lambda}}~~~\mbox{with}~~\nu_{\lambda}=x_{\lambda}\nu\simeq 0.567,
\end{equation}
while $x_{\lambda}=(\gamma/\nu)-x_D\simeq 0.902$.
Adopting the value $R_D\simeq 1.05$ \cite{26}
we find, using the values for $\Gamma_0$, $\xi_0$ and $\eta_0$ reported above, that a 
sound theoretical estimate of the amplitude $Q$ for our model is
\begin{equation}\label{qvalue}
Q=(2.8\pm0.4)\times 10^{-3}.
\end{equation}
\indent {\bf Finite-size scaling}.---~Since the background $\mathcal L_b(T)$ derives from
atomic length scales, it should vary little with $L$. However, the possibility of
significant finite-size effects on the critical part, $\Delta\mathcal L(T)$, must be
recognized and allowed for. Note, in particular, that although static properties may
(as here \cite{new9}) exhibit negligible finite-size deviations for the range of
($T-T_c$) and $L$ simulated (see Fig.~\ref{fig3}), the same need not be true for transport coefficients.
To tackle this problem
we write the finite-size scaling ansatz \cite{28,29,5} as
\begin{equation}\label{eq10}
\Delta\mathcal L/T^*\approx QW(y)\epsilon^{-\nu_{\lambda}},
\end{equation}
where $y=L/\xi$ while $W(y)$ is a finite-size scaling function
that must vary as $W_0y^{x_{\lambda}}[1+O(y^{1/\nu})]$ for small $y$, since
$\Delta\mathcal L(T_c;L)$ is finite for $L<\infty$ \cite{5,28,29}. For large $y$ one may
quite generally write
\begin{equation}\label{eq11}
W(y)=1+W_{\infty}e^{-ny}/y^{\psi}+~...~,
\end{equation}
where $W(\infty)$=$1$ is needed to reproduce (\ref{powerlawons}) when $L$$\rightarrow$$\infty$
while, for static properties in short-range systems, $n$ is a small integer \cite{5,28,29}.
However, for dynamic coefficients, where long-time tails, etc., may enter, 
one must be prepared for $n$\,=\,$0$ implying only an $L^{-\psi}$ decay of 
finite-size deviations; the exponent $\psi$ demands more detailed, currently
unavailable theory.\\
\indent To analyze the $\mathcal L(T;L)$ data a scaling plot of 
$\mathcal W_L(T)$\,$\equiv$\,$(\Delta\mathcal L/T^*)\epsilon^{\nu_{\lambda}}$ vs.~$y$ is
desirable: by (\ref{eq10}) and (\ref{eq11}) this should approach $Q$ for large $y$. But the background 
$\mathcal L_b(T)$, albeit slowly varying, is unknown! To meet this challenge, 
we introduce an effective background parameter $\mathcal L_b^{eff}$, and 
adjust it to optimize data collapse: See Fig.~\ref{fig3} which 
presents $\mathcal W_L(T)$ for three illustrative values of $\mathcal L_b^{eff}$ vs.~the bounded
variable $[y/(y_0+y)]^{x_{\lambda}}$ in which, purely for convenience, we have
set $y_0$\,=\,$7$. The optimal value, which serves as a rough estimate of $\mathcal L_b(T_c)$, 
proves to be $\mathcal L_b^{eff}$\,=\,$(3.3\pm0.8)$$\times$$10^{-3}$ \cite{new9}.
For this assignment we find that a good fit (see the dashed line
in Fig.~\ref{fig3}) is provided by $\mathcal W_L\simeq Q/[1+p_0/y(1+y^2/p_1^2)]^{x_{\lambda}}$
with $p_0$\,=\,$5.8\pm0.5$ and $p_1$\,$\simeq$\,$13.8$ while $Q$\,=\,($2.7$$\pm$$0.4$)\,$\times$\,$10^{-3}$. 
This estimate for $Q$ is in gratifying
agreement with the theoretical value reported in (\ref{qvalue}).\\
\indent The quantitative significance of the finite-size effects can be appreciated from Fig.~\ref{fig4} where
the scaling-function fit has been used to estimate $\mathcal L(T)$ for $N=2.56$$\times$$10^4$
and $N$$\rightarrow$$\infty$.
Note also that the fit for $\mathcal W_L(y)$ corresponds to $n=0$ and $\psi=3$ in (\ref{eq11}).
Further exploration suggests that if an ultimate exponential decay does arise
[if $n=1$ in (\ref{eq11})] it sets in only for $y=L/\xi\gg30$.\\
\indent {\bf In summary}.--- The extensive simulations we have performed for the transport properties
near the demixing point of our symmetric but otherwise not unrealistic binary
fluid model are, when appropriately analyzed with due attention to strong finite-size effects and
a background contribution, completely consistent with current theoretical 
predictions and the best available experimental data. Not only are exponent values
and the dynamic exponent relation (\ref{dynamicscaling}) respected but the
amplitude value (\ref{qvalue}) is also confirmed. While increased computer power and
more refined data analysis might eventually provide more stringent tests of theory,
such as the value of $R_D$, the necessary resources appear rather demanding.\\
\begin{figure}[htb]
\centering
\includegraphics*[width=0.44\textwidth]{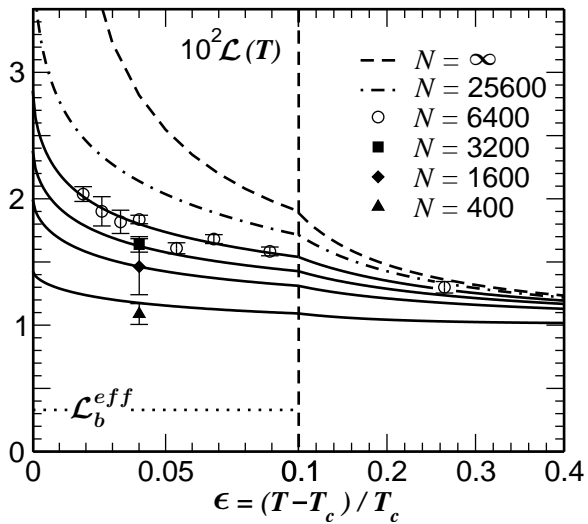}
\caption{\label{fig4}Variation of the Onsager coefficient with $T$ for
systems of increasing size based on the scaling function fit and 
optimal value of $\mathcal L_b^{eff}$ [$\simeq \mathcal L_b(T_c)$]:
see the dotted line. For $T$
within $1$\% of $T_c$, reliable estimates of $\mathcal L(T)$ would require
$N$\,$>$\,$3$\,$\times$\,$10^4$ and system sizes exceeding $30\sigma$.}
\end{figure}
\indent M.E.F. and S.K.D. are grateful to the National
Science Foundation for support
under Grant No. CHE $03$-$01101$ while S.K.D. also acknowledges 
the Deutsche Forschungsgemeinschaft
via Grant No.~Bi~$314/18$-$2$.
\vspace{-0.25cm}
%\newpage


\begin{thebibliography}{99}

\bibitem{1} J.V. Sengers and J.M.H. Levelt Sengers, Ann. Rev. Phys. Chem.
\textbf{37}, 189 (1986).

\bibitem{2} P. C. Hohenberg and B. I. Halperin, Rev. Mod. Phys.
\textbf{49}, 435 (1977).

\bibitem{3} K. Binder and E. Luijten, Phys. Repts. \textbf{344}, 179 (2001).

\bibitem{4} G. Orkoulas, M.E. Fisher, and A. Z. Panagiotopoulos, Phys. Rev. E
\textbf{63}, 051507 (2001).
E. Luijten {\it et al.}, 
Phys. Rev. Lett. \textbf{88}, 185701 (2002).

\bibitem{5} Y. C. Kim and M. E. Fisher,
Phys. Rev. E \textbf{68}, 041506 (2003); Phys. Rev. Lett.
\textbf{92}, 185703 (2004).

\bibitem{7} K. Jagannathan and A. Yethiraj, Phys. Rev. Lett. \textbf{93},
015701 (2004); J. Chem. Phys. \textbf{122}, 244506 (2005);
Phys. Rev. Lett. \textbf{94}, 069602 (2005).

\bibitem{nn7} More recently, A. Chen {\it et al.}, Phys. Rev. Lett.
\textbf{95}, 255701 (2005), simulated the thermal equilibration of
a single-component fluid at criticality and found a result in 
accord with theory \cite{2}.

\bibitem{8} J. V. Sengers and M. R. Moldover, Phys. Rev. Lett.
\textbf{94}, 069601 (2005).

\bibitem{new9} S.K. Das, J. Horbach, K. Binder, M.E. Fisher, and J.V. Sengers, 
cond-mat/0603587 (2006).

\bibitem{9} A. Sariban and K. Binder, J. Chem. Phys. \textbf{86}, 5859
(1987); 
S. K. Das, J. Horbach, and K. Binder, J. Chem. Phys.
\textbf{119}, 1547 (2003). 

\bibitem{12} D. P. Landau and K. Binder, \textit{A Guide to Monte Carlo
Simulations in Statistical Physics}, 2nd ed. (Cambridge Univ.
Press, Cambridge, 2005).

\bibitem{13} K. Binder, Z. Phys. B \textbf{43}, 119 (1981).

\bibitem{14} See, e.g., J. Zinn-Justin, Phys. Repts. \textbf{344}, 159 (2001).

\bibitem{15} M. P. Allen and D. J. Tildesley, \textit{Computer Simulations
of Liquids} (Clarendon Press, Oxford, 1987).

%\bibitem{16} K. Binder and G. Ciccotti (eds) \textit{Monte Carlo and
%Molecular Dynamics of Condensed Matter Systems} (Italian Physical
%Society, Bologna, 1996).

\bibitem{17} R. Kutner, K. Binder, and K.W. Kehr, Phys. Rev. B \textbf{26}, 2967 (1982).

\bibitem{18} A.N. Drozdov and S.C. Tucker,
J. Chem. Phys. \textbf{114}, 4912 (2001); \textbf{116}, 6381 (2002).

\bibitem{19} K.R. Harris, J. Chem. Phys. \textbf{116}, 6379 (2002).

\bibitem{20} H. Hao {\it et al.},
Phys. Rev. E \textbf{71}, 021201 (2005).

\bibitem{21} R.F. Berg {\it et al.}, 
Phys. Rev. Lett. \textbf{82}, 920 (1999).
%; Phys. Rev. E \textbf{60}, 4079 (1999).

\bibitem{22} J.-P. Hansen and I.R. McDonald, {\it Theory of Simple Liquids} 
(Academic, London, 1986).

\bibitem{23} H.C. Burstyn and J.V. Sengers, Phys. Rev. Lett. \textbf{45},
259 (1980); Phys. Rev. A \textbf{25}, 448 (1982).

%\bibitem{24} K. Kawasaki, in \textit{Phase Transitions and Critical
%Phenomena, Vol 5A}, edited by C. Domb and M. S. Green (Academic,
%New York, 1976), p. 165.

\bibitem{25} J.V. Sengers, Int. J. Thermophys. \textbf{6}, 203 (1985).

\bibitem{26} J. Luettmer-Strathmann {\it et al.},
J. Chem. Phys. \textbf{103}, 7482 (1995);
J. Luettmer-Strathmann and J. V. Sengers, J. Chem.
Phys. \textbf{104}, 3026 (1996).

\bibitem{27} J.V. Sengers and P.H. Keyes, Phys. Rev. Lett.
\textbf{26}, 70 (1971);
H.L. Swinney and D.L. Henry, Phys. Rev. A \textbf{8},
2586 (1973).

\bibitem{28} M. E. Fisher, in \textit{Critical Phenomena}, edited by M. S.
Green (Academic Press, London, 1971) p. 1.

\bibitem{29} V. Privman, ed., \textit{ Finite Size Scaling and Numerical
Simulation of Statistical Systems} (World Scientific, Singapore,
1990).

\end{thebibliography}
\end{document}